\documentclass[9pt,twocolumn,twoside]{opticajnl}
\journal{opticajournal} 
\setboolean{shortarticle}{true}
\setboolean{displaycopyright}{true} 

\usepackage{upgreek}
\usepackage{lineno}
\title{Phase stabilization for long baseline interferometry of incoherent optical sources}

\author[1,*]{Joshua J. Collier}
\author[1,2]{David R. Gozzard}
\author[1]{John S. Wallis}
\author[1,2]{Benjamin P. Dix-Matthews}

\affil[1]{International Centre for Radio Astronomy Research, The University of Western Australia, Crawley WA 6009, Australia}
\affil[2]{ARC Centre of Excellence for Engineered Quantum Systems (EQUS), Department of Physics, The University of Western Australia, Crawley WA 6009, Australia}
\affil[*]{joshua.collier@research.uwa.edu.au}

\begin{abstract}
The maximum baseline, and therefore resolution, of optical astronomical interferometers is limited by attenuation and phase noise within the optical path between the apertures and beam combiner, as well as the practical challenges of constructing optical delay lines more than a few hundred meters in length. We implement off-band phase stabilization on two fiber optic links of 85~km, creating a total baseline of 170~km. We show that the system is able to effectively phase stabilize signals from an incoherent pseudo-thermal source with a bandwidth of 11.2~nm. We are able to reduce the phase noise 
by 4-5 orders of magnitude between 1 and 100~Hz such that we could resolve an applied phase difference of 0.16~cycles per second with continuous measurement. We show that, with phase stabilization active, the interferometer is able to recover both first-order and second-order photon correlations. These results demonstrate the feasibility of this technique for long-baseline optical and quantum astronomical interferometers. The present results are limited by chromatic dispersion within the fiber, which can be mitigated using dispersion compensating modules.
\end{abstract}

\begin{document}

\maketitle

It is well established that the resolution of our terrestrial telescopes, and thus our observations of the universe, are classically limited by the size of the primary aperture, and the wavelength of light being observed \cite{Rayleigh1879}. Techniques such as aperture synthesis \cite{Thompson2017} have allowed us to increase resolution using long baselines between two or more apertures. These interferometers are still classically limited by the Rayleigh criterion, a heuristic set by the merging of the Airy discs of multiple sources due to the finite point-spread function (PSF) of the imaging system. Quantum metrology has become an important field in overcoming this Rayleigh limit \cite{Tsang2016, Tham2017,tsang2019resolving}. By making quantum-optimal measurements of the available photons these quantum metrology methods allow us to extract more information from our observations. Modal decomposition of the beam, like spatial mode demultiplexing (SPADE) \cite{tsang2019resolving,Wallis2025,gozzard2025super} allows us to approach the quantum limit for source separation estimation \cite{Ang2017}. Hanbury Brown-Twiss intensity interferometry has also been used \cite{Thiel2007, Oppel2012, Liu2025, Kim2025} for sub-Rayleigh source separation estimation. This technique has been expanded to two-photon amplitude interferometry to extract more information from each photon \cite{Stankus2022, Crawford2023}. An interferometric method, that is equivalently quantum limited as SPADE, was proposed by Pearce et al. \cite{Pearce2017} and experimentally demonstrated in the works of Howard et al. \cite{Howard2019} and Zanforlin et al. \cite{Zanforlin2022}. This method has become known as quantum optical interferometry. Using single photon correlations, we can overcome the Rayleigh limit to achieve `super-resolution' --- resolution better than the classical Rayleigh or diffraction limit. 
These works were performed over small baselines of less than 5~cm, limited by phase noise in the interferometer. Phase noise is a major limitation on building long-baseline classical or quantum optical interferometers. 

\begin{figure*}[h]
    \centering
    \includegraphics[width=0.91\linewidth]{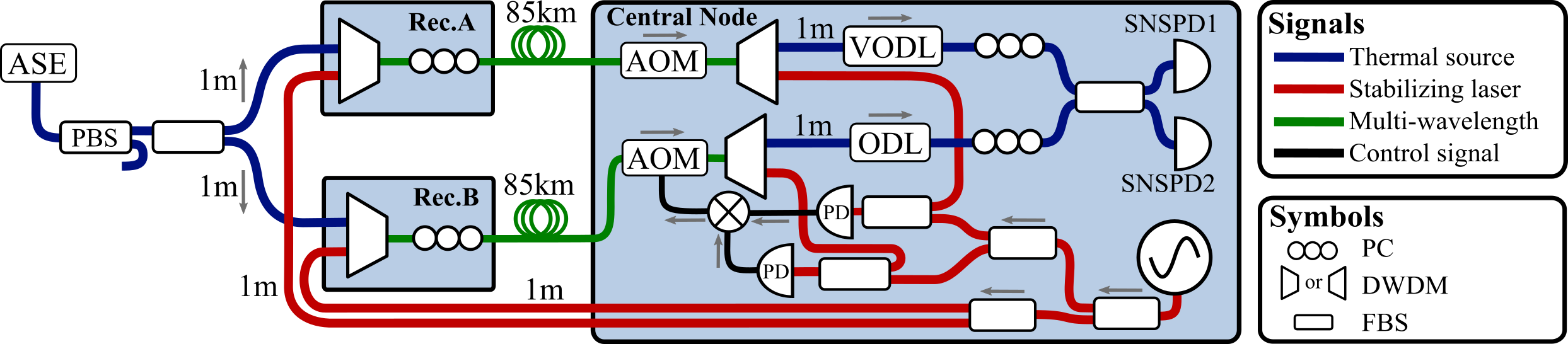}
    \caption{Functional System Diagram. The system under test placed both receiver A and B next to each other and the two 85 km fibers are fibers that run next to each other around the city of Perth (Boorloo), Western Australia, Australia. The blue fibers are our thermal source collected from the receivers, the red fibers are the stabilizing laser, from an X15 at 1552.51nm, and the green fiber is where the signals are combined (multi-wavelength). DWDM: dense wavelength division multiplexor (CH31 separating), PD: photodetector, SNSPD: superconducting nanowire single-photon detector, (V)ODL: (variable) optical delay line, AOM: acousto-optic modulator, PC: polarization controller, FBS: fiberized beam splitter, ASE: amplified spontaneous emission, PBS: polarizing beam splitter}.
    \label{fig:system_diagram}
\end{figure*}

The longest existing optical interferometer is the Center for High Angular Resolution Astronomy (CHARA) array, with a maximum baseline of 330~m \cite{Brummelaar2005}. CHARA uses a large free-space delay line beneath the telescopes. From an engineering perspective, building these large delay lines becomes significantly harder as the baselines increase. A transition to fiber based delay lines will be necessary to realize interferometers with baselines greater than a few kilometers long \cite{koehler2024integrating}. Fiber optic systems couple significantly more phase noise from their environment and this reduces our interferometer's visibility. Active phase stabilization of the interferometer arms is required to recover a usable interferometric measurement, and such systems have been used on telescopes such as the Square Kilometre Array \cite{schediwy2019mid}. However, compared with these previous systems, phase stabilization for optical interferometry imposes additional challenges including the need to provide effective stabilization across wide (100$+$~nm) optical bandwidths, and avoiding the relatively bright phase stabilization signals degrading the measurement of faint astronomical signals. This latter point is particularly important for quantum interferometers where noise resulting from the bright probe signal can severely degrade the single photon measurements. Actively phase-stabilized frequency reference distribution systems have been used in radio-frequency astronomical interferometers for many years \cite{cliche2006precision,schediwy2019mid}, and phase-stabilized quantum interferometers spanning hundreds of kilometers have been implemented for quantum communications in recent years \cite{Clivati2022, Pittaluga2021}. Here, we extend the technique to stabilize broadband pseudo-thermal signals emulating star light. In this paper we demonstrate phase stabilization of star-like signals over a 170~km baseline. 



We have prepared both a pseudo-thermal source and an interferometric receiver as shown in Figure~\ref{fig:system_diagram}. Our thermal source is an unseeded erbium-doped fiber amplifier (EDFA) which is temporally incoherent with a coherence length of 65~$\mu$m and produces unpolarized light with a central wavelength of 1530~nm across a linewidth of 11.2~nm due to the amplified spontaneous emission (ASE) noise. This then passes through a polarizing beam splitter so that we have a single polarization state, and passed through a series of attenuators to reduce the initial power to -65~dBm. We couple the source into our interferometer using a fiber beam splitter. 


To build the interferometer, we have two receivers that couple light from the source (referred to as the thermal signal). We use an X15 as a stabilizing laser (referred to as the stabilizing signal) with a central wavelength of 1552.51~nm, and a linewidth of 100~Hz. We multiplex the thermal signal with -40~dBm of the stabilizing signal using dense wavelength division multiplexors (DWDMs). Once the channels are combined, these signals travel through 85~km of fiber optic cable to emulate distant receivers (referred to as an arm). Each arm then has a polarization controller, and an acousto-optic modulator (AOM) that shifts the frequency by 85~MHz. A full-scale interferometer of this type would require the outgoing stabilizing signal to travel a distance out to the receiver equal to the return path as in \cite{Clivati2022, Pittaluga2021}. We only had two 85~km links available. 
However, we do not expect this to make a significant difference to the present results \cite{Clivati2022,Pittaluga2021}.

The stabilizing laser is then demultiplexed from each arm with another pair of DWDMs. We measure the optical phase noise from the beatnote between the returning stabilizing signal and a local -20~dBm signal from the stabilizing laser \cite{Haus2000}. The phase noise signals from each arm are mixed together electronically, producing a DC error signal proportional to the difference in phase of the two arms. We use this error signal to drive the phase of one arm to follow the other in a closed loop via frequency modulation of one of the AOMs. The broadband thermal signal from each arm travels through $\approx$50~mm variable optical delay lines that can be used for introducing a fixed phase difference, as well as fine path length matching. The thermal signal from both arms is then interfered on a fiberized beam splitter and both outputs are measured by superconducting nanowire single-photon detectors (SNSPDs). A first order interference measure can be taken directly from the fringes observed on the the SNSPDs as a phase difference between the arms is induced by changing the relative path length of the arms with one of the optical delay lines. A Liquid Instruments Moku:Pro is used for the the control system. Interference measurements were taken using the counts of the SNSPDs with a Swabian Time Tagger 20. This device has 34~ps RMS timing jitter.


\begin{figure}[htb!]
    \centering
    \includegraphics[width=0.9\linewidth, trim=0.45cm 0.5cm 0.5cm 0.45cm, clip]{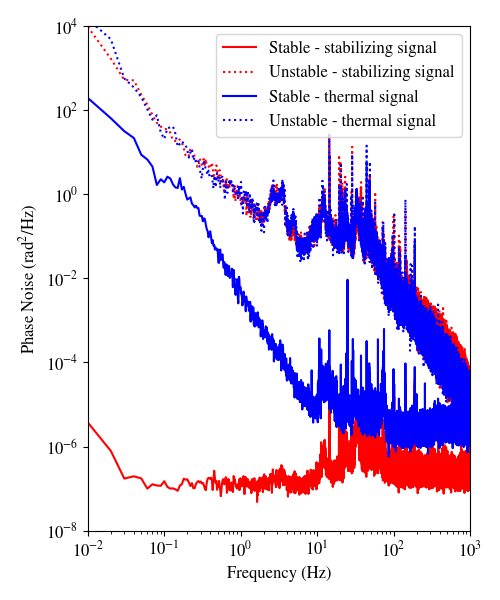}
    \caption{Power spectral density (PSD) of link with and without phase stabilization. Red curves are the phase difference measured between the two stabilizing signals without phase stabilization system (dotted) and with phase stabilization (solid). The blue curves are the phase of one of the thermal signals without phase stabilization (dotted) and with phase stabilization (solid). The thermal signal here is from a coherent source (a laser with 10~kHz linewidth).}
    \label{fig:psd_phasestab}
\end{figure}

In Figure \ref{fig:psd_phasestab} we have phase noise measurements from both the stabilizing signal and the thermal signal. The stabilizing signal data were taken with a phasemeter on both arms tracking the 85~MHz signal and calculating the noise of the difference between the arms. This shows how effective the stabilization is at correcting noise on the thermal signal at a fixed wavelength (1554~nm with 10~kHz linewidth). It is shown that there is up to 9 orders of magnitude phase correction for our in loop measurement (our stable stabilizing signal). The out of loop measurement (our stable thermal signal) sees only 4-5 order of magnitude phase noise reduction between 1 and 100~Hz. We note that these 85~km fibers are traveling through the same cable, and thus some of the environmental noise is correlated.

In Figure \ref{fig:psd_phasestab} the thermal phase noise is measured by standard fibre photodetectors in place of the SNSPDs, as well as an additional AOM before one of these receiving detectors in order to produce a heterodyne tone for the phasemeter to measure. The "thermal signal" in this instance is a coherent source at 1554~nm so that we could perform this coherent detection. We can see the stabilization below 10~Hz of the thermal signal kick up, this is due to the difference in wavelength of the thermal and stabilizing signal. The stabilizing laser is used to stabilize the two arms relative to each other, to the limit
\begin{equation}
    S_{min}(f)=\frac{(\lambda_s-\lambda_t)^2}{\lambda_s^2}\frac{lL}{f^2}
    \label{eq:stab_limit}
\end{equation}
from Bertaina et al. \cite{Bertaina2024}, where $l$ is a coefficient of noise in fiber, $L$ is the length of the arms, $\lambda_s$ is the stabilizing laser wavelength, and $\lambda_t$ is the thermal source wavelength. This limit comes from the offset of the wavelengths when corrected via a phase delay actuator instead of a group delay actuator. This limit is not directly applicable here as a result of the correlated noise between the fibers due to the fibers sharing a similar physical space. However, in a real-world deployment, as in Bertaina et al. \cite{Bertaina2024}, the noise in the two fibers would be uncorrelated, producing the stability limit given by Equation~\ref{eq:stab_limit}. We also expect a small amount of phase noise to be induced from the out of loop fiber. Given that our experiment is conducted in a temperature stable lab, and these fiber lengths have been limited, this effect should be negligible.

\begin{figure}[htb!]
    \centering
    \includegraphics[width=0.98\linewidth, trim=0.2cm 0.08cm 0.05cm 0.1cm, clip]{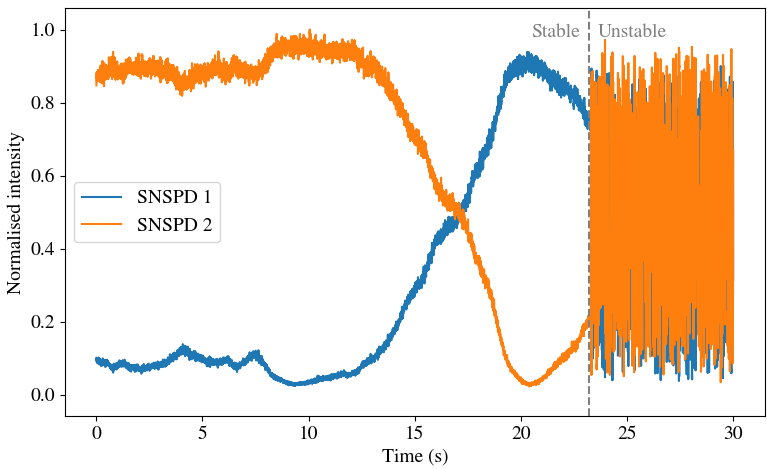}
    \caption{Measured intensity over time without changing the phase offset. 
    Measuring the number of counts in both thermal signals (with a coherent source) over time to observe the drift in the system.} 
    \label{fig:countsOverTimeDrift}
\end{figure}

\begin{figure}[htb!]
    \centering
    \includegraphics[width=0.95\linewidth]{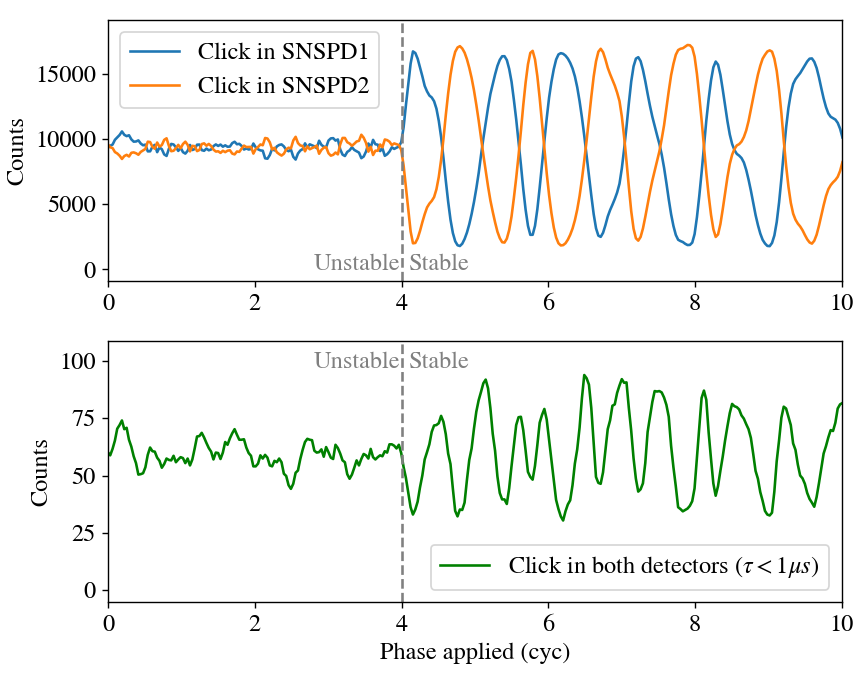} 
    \caption{Photon counts and correlations over time with a coherent source. Top plot is first order interference [0,1] and [1,0] states. Bottom plot is second order interference [1,1] state. The counts are per bucket, where buckets are 200~ms long. Here, $\uptau$ is the time period over which we consider a coincidence event.
    } 
    \label{fig:coherent_first_second_order}
\end{figure}

\begin{figure}[htb!]
    \centering
    \includegraphics[width=0.95\linewidth, trim=0.2cm 0.08cm 0.05cm 0.1cm, clip]{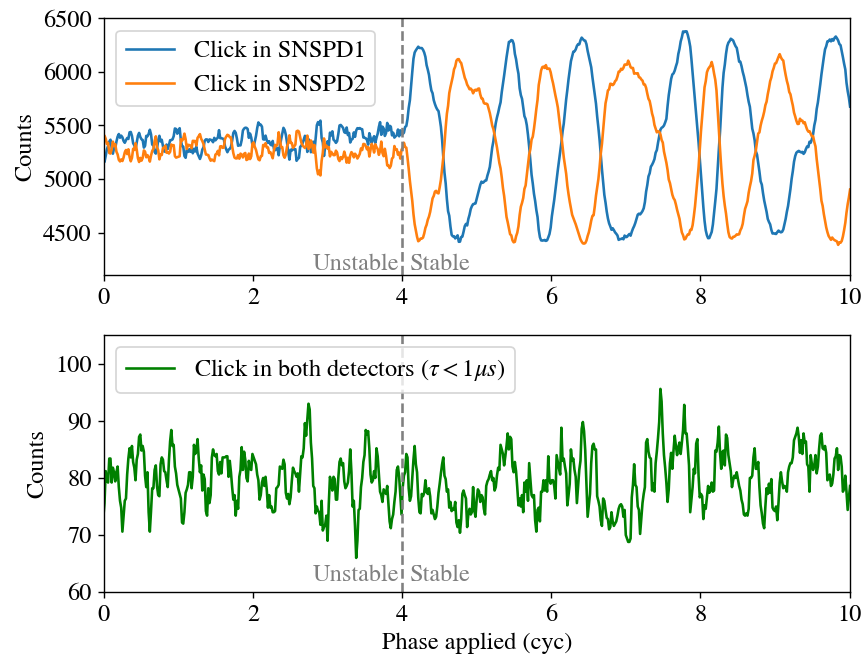}
    \caption{Photon counts and correlations over time with an incoherent source. Top plot is first order interference [0,1] and [1,0] states. Bottom plot is second order interference [1,1] state.  The counts are per bucket, where buckets are 200~ms long. Here, $\uptau$ is the time period over which we consider a coincidence event.
    }
    \label{fig:incoherent_first_second_order}
\end{figure}

There is significantly more noise reduction in the stabilizing signal in-loop measurement. The noise in the thermal signal could be further reduced by integrate over significantly longer periods with a more involved stabilization system (i.e. using multiple stabilizing beams to determine the wavelength dependence of the noise to correct for the wavelength band of the thermal signal or employing group-delay actuators instead of the AOM).
Despite this, there is a significant reduction in phase noise in the interference of the thermal signals, emphasized in the time domain. Figures \ref{fig:countsOverTimeDrift}, \ref{fig:coherent_first_second_order}, and \ref{fig:incoherent_first_second_order} demonstrate this time domain behavior.

In Figure~\ref{fig:countsOverTimeDrift} we can see the phase drift over time without applying a phase offset between the two thermal signals both with and without the phase stabilization. While the phase stabilization system is active, we see less than one cycle phase shift over the 25 seconds. Without the phase stabilization system the signal is completely unrecoverable at 100~Hz sampling as shown here. This is further demonstrated in Figures \ref{fig:coherent_first_second_order} and \ref{fig:incoherent_first_second_order}. 

We see interference for our coherent source (a laser at 1554~nm, with a bandwidth of 0.1~fm/10~kHz) in Figure \ref{fig:coherent_first_second_order}, and our incoherent source (an unseeded EDFA, with a bandwidth of 11.2~nm/1.4~THz) in Figure \ref{fig:incoherent_first_second_order}. These figures show both first order interference (where either SNSPD1 or SNSPD2 receive a photon), as well as second order interference (where SNSPD1 and SNSPD2 receive a photon within a given time period).


In Figure \ref{fig:coherent_first_second_order} we can see the sinusoidal interference fringe produced by a ramped phase offset between the thermal signals both with and without the phase stabilization. Here we swept the phase at 0.16 cycles per second with the variable optical delay line. Both the first order and second order fringes are recovered while phase stabilization is active. Recovering the second order interference fringe here is important for application to quantum interferometry \cite{Howard2019}. We use 200~ms binning to optimally balance high SNR while still recovering our interference fringe.

Figure \ref{fig:incoherent_first_second_order} shows the same test using our incoherent source. Here we have swept the phase at 0.16 cycles per second. With the phase stabilization the first order fringe is recovered with reduced visibility due to chromatic dispersion in the fiber. A second order interference is not recovered in this instance, as it is significantly degraded by the chromatic dispersion. 

We were able to recover full depth fringes (100\%) from the unseeded EDFA at the center of the interference fringe before adding the 85~km fibers to each arm. Due to chromatic dispersion the maximum visibility at the center of the first-order interference fringe dropped to 25\%. Since chromatic dispersion is a linear effect, with a significant amount of dispersion compensation, this should be easily mitigated with dispersion compensating fiber or a fiber Bragg grating \cite{Guenther2018, Gul2023}. 

We have demonstrated the feasibility of wideband phase stabilization for optical and quantum astronomical interferometers. Our phase stabilization system allows us to resolve fringes of both coherent and incoherent sources with 0.16~cycles per second phase difference induced, meaning the interference can be measured over significantly longer integration times than the unstabilized interferometer. This will allow for greatly increased the signal-to-noise ratio, and higher loss / longer interferometer arms. Second-order correlation photon fringes are also obtained, showing that the wideband stabilization is also effective for application in quantum interferometers. Future works should investigate applying dispersion compensation to recover full visibility from a wide-band source, thereby fully demonstrating the potential of this interferometer for long baseline astronomical optical interferometry. The use of group delay actuators for stabilization, or the addition of a periodic global phase reset, will be needed to increase the achievable integration times. 

The removal of phase noise accrued on the path the light from the two receivers has traveled allows us to produce a fiber-based interferometer where the receivers are both 85 km away from the central node. With a diffraction-limited system, a 170~km baseline optical interferometer at 1550~nm could resolve, according to the Rayleigh criterion, an angle of approximately 2.3~micro-arcseconds, 
more than an order of magnitude greater than the record 20~micro-arcsecond resolution achieved by the globe-spanning Event Horizon Telescope \cite{akiyama2019first}. This is highly sought after for myriad astronomical measurements such as the detection and characterization of exoplanets, investigation of star and planet formation, and the study of extreme spacetime curvature near the event horizon of black holes. This work demonstrates a practical system that can be used in building this long baseline interferometer.

\begin{backmatter}
\bmsection{Funding} Australian Research Council (project ID CE17010009, project ID DE240100587), Air Force Office of Scientific Research (project ID FA2386-23-1-4081).
\bmsection{Acknowledgements} J.J.C and J.S.W are supported by Australian Government Research Training Program Scholarships and top-up scholarships funded by the Government of Western Australia. This material is based upon work supported by the Air Force Office of Scientific Research under award number FA2386-23-1-4081. The authors thank the ICRAR-UWA Astrophotonics group for useful conversations and assistance, and thanks also to AARNet for the provision of light-level access to their fiber network infrastructure.
\bmsection{Disclosures} The authors declare no conflicts of interest.
\bmsection{Data availability} Data underlying the results presented in this paper are not publicly available at this time but may be obtained from the authors upon reasonable request.
\end{backmatter}

\bibliography{references}

\bibliographyfullrefs{references}

\end{document}